# Spin-orbit torque based physical unclonable function

G. Finocchio,[1,*] T. Moriyama,[2] R. De Rose,[3] G. Siracusano,[4] M. Lanuzza,[3] V. Puliafito,[1] S. Chiappini,[5] F. Crupi,[3] Z. Zeng,[6] T. Ono,[2] M. Carpentieri[7,*]

[1]Department of Mathematical and Computer Sciences, Physical Sciences and Earth Sciences, University of Messina, Messina 98166, Italy

[2]Institute for Chemical Research, Kyoto University, Uji, Kyoto 611-0011, Japan

[3]Department of Computer Engineering, Modeling, Electronics and Systems Engineering, University of Calabria, I-87036 Rende, Italy

[4]Department of Electric, Electronic and Computer Engineering, University of Catania, I-95125 Catania, Italy

[5]Istituto Nazionale di Geofisica e Vulcanologia, Via di Vigna Murata 605, I-00143 Roma, Italy

[6]Key Laboratory of Multifunctional Nanomaterials and Smart Systems, Suzhou Institute of Nano-Tech and Nano-Bionics, CAS, Suzhou, Jiangsu 215123, People's Republic of China

[7]Department of Electrical and Information Engineering, Polytechnic of Bari, Bari 70125, Italy

[*]Corresponding authors: gfinocchio@unime.it, mario.carpentieri@poliba.it

**Abstract**

This paper introduces the concept of spin-orbit-torque-MRAM (SOT-MRAM) based physical unclonable function (PUF). The secret of the PUF is stored into a random state of a matrix of perpendicular SOT-MRAMs. Here, we show experimentally and with micromagnetic simulations that this random state is driven by the intrinsic nonlinear dynamics of the free layer of the memory excited by the SOT. In detail, a large enough current drives the magnetization along an in-plane direction. Once the current is removed, the in-plane magnetic state becomes unstable evolving towards one of the two perpendicular stable configurations randomly. In addition, an hybrid CMOS/spintronics model is used to evaluate the electrical characteristics of a PUF realized with an array of 16×16 SOT-MRAM cells. Beyond robustness against voltage and temperature variations, hardware authentication based on this PUF scheme has additional advantages over other PUF technologies such as non-volatility (no power consumption in standby mode), reconfigurability (the secret can be rewritten), and scalability. We believe that this work is a step forward the design of spintronic devices for application in security.

# I. Introduction

In the "Internet of Things" (IoT) era the security is becoming a crucial aspect and a path to enhance the security of a physical device is the hardware authentication. To provide such a secure authentication procedure, hardware cryptographic operations (such as digital signatures or encryption) having a secret key in a nonvolatile electrically erasable programmable read-only memory or battery backed static random-access memory are used.[1] However, those approaches are expensive in terms of both area occupancy and power consumption and a challenging solution should be to have hardware authentication based on a memory technology integrated within the device itself. The use of Physical Unclonable Functions (PUFs) is a direction to face this challenge promising to enhance the security of the device at a minimal additional hardware cost.[2]

PUFs are innovative primitives that derive a chip-unique challenge-response mechanism by typically exploiting the randomness due to manufacturing process variations. In other words, the challenge-response mechanism converts the unique physical state of the PUF into a digital input-output data. The two main subtypes of PUFs are "weak PUFs" and "strong PUFs": the former store the secret in vulnerable hardware while the latter have complex challenge-response behavior from the physical disorder characterizing the PUF itself.[3] The most popular implementation of weak PUFs are the SRAM PUFs, while one of the first implementations of a strong PUF is based on optical scattering.[2] Recently, the major semiconductor foundries have integrated spintronics with CMOS technology in their processes, in particular for the fabrication of STT-MRAMs (spin-transfer-torque magnetic random access memories).[4],[5] This opens a large number of opportunities and in particular we wish to look at possible implementations of PUF with spintronic technology.[6]

The commercial PUFs have several problems, which can affect their performance and reliability, related to environmental and/or operative variations that have to be addressed in the design and test phases, such as temperature and supply voltage dependent response, and electromagnetic interference. Spintronic technology thanks to the development of reliable magnetic tunnel junctions (MTJs) can be used to fix some of those problems. For examples, MRAMs have been already proposed for PUF development, where the secret can be derived from magnetic textures, such as domain wall,[7] or in STT-MRAMs by switching probability[8]. Fig. 1(a) shows the device to implement the latter STT-MRAM based PUF. First it is evaluated the switching probability of the free layer (FL) as a function of the current pulses of a certain duration (the voltage is supplied at the terminals 'A' and 'B'). Once determined the current value and duration at which the switching probability is 50%, this is applied to a matrix of STT-MRAMs, where all the cells have been set at the same initial state, e.g. FL parallel to the PL (pinned layer), thus driving a random state that generates the secret key of the PUF. STT-MRAMs can be designed with an energy barrier larger than $60k_BT$ making the secret thermally very stable.[9] In addition, the reading signal is insensitive to voltage fluctuations and the resistance states exhibit a reduced drift for a wide range of temperature near room temperature (a typical tunneling magneto resistive (TMR) ratio is of the order of 100%)[10] that make the states intrinsically very distinguishable. On the other hand, the derivation of the secret is the main problem being related to the switching probability that is very sensitive to temperature changes and device-to-device geometrical variations.

To address this aspect, the idea developed in this work is to use the spin-orbit-torque MRAMs (SOT-MRAMs) as a building block for the PUF. An SOT-MRAMs is a three terminal device where the read and write operations are separated (see Fig. 1(b)). The information is written with the SOT originated by a current flowing in a heavy metal (HM) (Pt, Ta, W, Ir) mainly due to the spin-Hall effect ($J_{SHE}$), while the information is read via the resistance of the MTJ built on top of the HM. The main difference of an SOT-MRAM based PUF compared with the STT-MRAM counterpart is the writing mechanism, hence all the benefits from the reading are the same.

The starting point for the development of SOT-MRAMs is due to the pioneering paper by Miron et al.[11] in 2011, where it is demonstrated the switching of a single perpendicular ferromagnet coupled with a HM having a large spin-orbit coupling, Pt/Co in that paper, driven by a current flowing into the HM. A similar result was achieved by Liu et al.[12] in a three-terminal MTJ having the FL at the interface with the HM. Micromagnetic simulations performed to study the magnetization dynamics in those devices show that for perpendicular ferromagnets the switching occurs via a nucleation process and that a stochastic switching is also possible.[13] Basically, at zero field the current drives the magnetization at an intermediate state characterized by a near zero net out-of-plane component of the magnetization, that can then evolve to an out-of-plane state when the bias current is switched off. Differently here, we demonstrate that an intermediate uniform in-plane state can be driven by a proper combination of applied current and external field, and that also for this configuration once the current is removed, the intermediate state becomes unstable and the magnetization evolves randomly towards the positive or negative uniform out-of-plane state. A schematic of this behavior is shown in Fig. 1(c). While those intermediate states have been used to demonstrate clocking driven by current in magnetic logic,[14],[15] here we show how those can be used as a mechanism to generate the secret of an SOT-MRAM based PUF. In other words, this approach does not imply the intrinsic random process variations of the manufacturing process, but the secret key is introduced by the random time domain evolution of an unstable magnetic state towards two different other states that code the bit "1" and "0". Micromagnetic simulations are employed to reproduce the experimental data and the statistical properties are then used to design and simulate an SOT-MRAM based PUF with an hybrid CMOS/spintronics model into Cadence$^{TM}$ circuit design tool.[16]

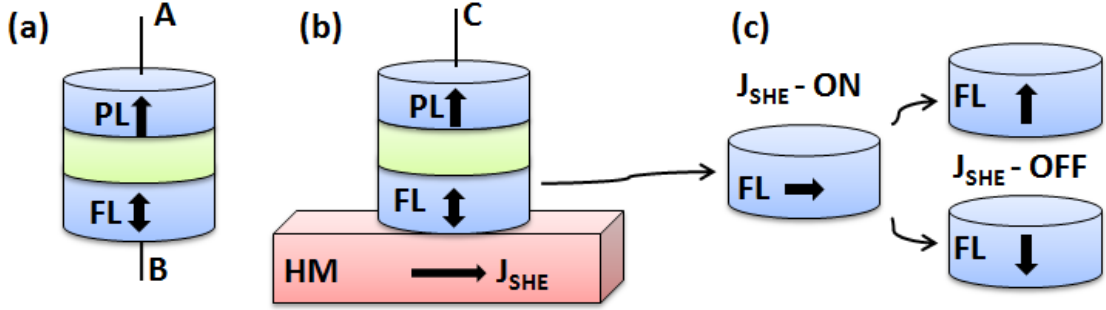

**Figure 1 a.** A schematic of a two terminals STT-MRAM with the indication of the free layer (FL) and pinned layer (PL). A current flowing through the terminals A and B manipulates the magnetic state of the FL. For those devices, random bits can be generated with current pulses that give rise to a 50% of the switching probability. **b.** A schematic of a three terminal SOT-MRAM. The writing and reading currents are separated, the current flowing from terminal C into the pillar is used to read the resistance of the MTJ while the current applied to the heavy metal ($J_{SHE}$) manipulates the magnetic state of the FL. **c.** In the SOT-MRAM having a perpendicular FL, the $J_{SHE}$, which has a spin polarization along the in-plane direction perpendicular to its flowing direction, can drive the FL magnetization in the in-plane configuration. Once the $J_{SHE}$ is switched off, the in-plane magnetization can evolve to both positive or negative out of plane directions with equal probability. This mechanism can be used to generate random bits in SOT-MRAMs which can be used as a secret key in an SOT-MRAM based PUF.

## II. Device and measurements.

Magnetic multilayers W(6)/Co$_{40}$Fe$_{40}$B$_{20}$(1)/MgO(1.6)/Ta(1), thickness in nm, are prepared by magnetron sputtering. The multilayers are then patterned into the device shown in Fig. 2(a) where the ferromagnetic dot has a circular shape with a diameter of 1μm on a 5μm-wide HM strip of *W* by e-beam lithography and ion-milling technique. The Co$_{40}$Fe$_{40}$B$_{20}$ is magnetized perpendicular to the sample plane due to the strong perpendicular magnetic anisotropy (PMA) originating from the Co$_{40}$Fe$_{40}$B$_{20}$/MgO interface[10]. The perpendicular magnetic anisotropy is larger than the out-of-plane demagnetizing field in order to set the out-of-plane direction as the easy axis of the magnetization. Typical magneto-optical Kerr microscope image of the device is shown in Fig. 2(b). The circled part is the location of the dot and it is where the contrast changes up on the magnetization switching. The electric current flowing in the W stripe invokes the spin current injection into the ferromagnetic dots due to the spin-Hall effect. The spin current exerts a torque resulting in a manipulation of the FL magnetization.

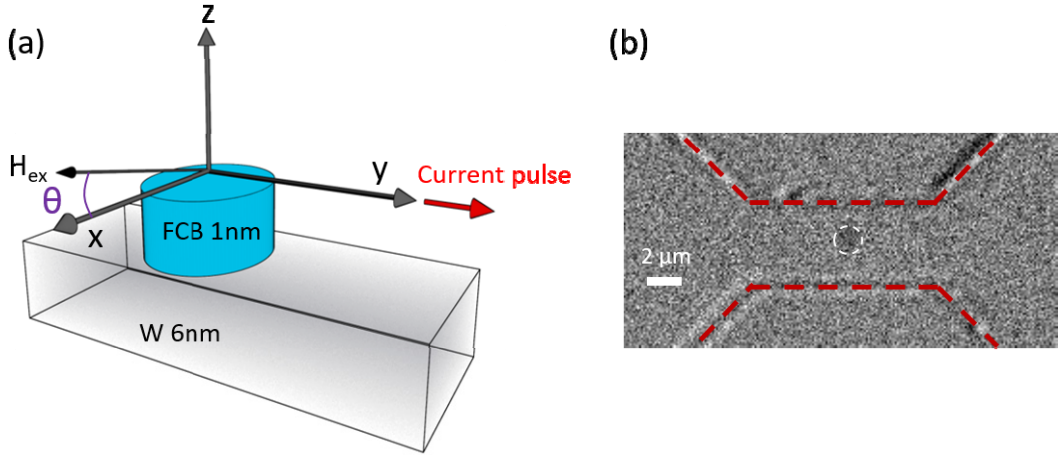

**Figure 2. a.** A schematic illustration of the experimental device with the indication of the coordinate reference system. A dot composed of $Co_{40}Fe_{40}B_{20}$ (FCB) (1 nm)/MgO (1.6 nm)/Ta (1 nm) is build on top a tungsten layer. During the measurements, an external field is applied along the x-axis with a tilting angle $\theta$ with respect to the sample plane. A current pulse is applied along the y-direction to the W stripe giving rise to a spin-orbit-torque at the interface with FCB. **b.** Magneto-optical Kerr microscope image showing a FCB dot (the darker part indicated by the white circle) on the W stripe (indicated by the red dotted lines).

**III Single dot measurements.**

The first set of measurements is performed on a single dot. A current pulse of 4 mA (voltage 8 V) is applied for a duration of 15 ns together with an external magnetic field of 40.2 mT applied along the x-direction (same direction of the spin-polarization) with a tilting angle $\theta$ with respect to the sample plane. The field amplitude is chosen in order to be much larger than the in-plane field like torque and the Oersted field originated by the current flowing in the HM, estimated around 0.5 mT for 4 mA by Biot-Savart law. The current drives the magnetization out of the equilibrium. When the current pulse is removed, the magnetization goes back to the perpendicular direction, both upward or downward, as indicated by optical imaging using magneto-optical Kerr effect. The switching probability (transition from in-plane to out-of-plane states) characterizing the final state, achieved for 500 current pulses, has been studied for different tilting angle ($\theta$ =-1, -1.6, -1.8 and -2 degrees) as displayed in Figs. 3(a)-(d), which describe the MOKE contrast measured after each current pulse. Those data are summarized in the histograms of Figs. 3(e)-(h) and in the probability phase diagrams of Figs. 3(i)-(l), with the indication of the switching down-to-up, up-to-down, down-to-down, and up-to-up, where up (down) is referred to the upward (downward) state of the magnetization. The switching probabilities are sensitive to the tilting angle of the magnetic field originating a strong variation in one degree. However, the switching probability to have upward or downward final state can be very close to 50% by a fine tuning of this angle (-1.8 degree), value that in ideal samples should be zero. We argue that it is non zero in the experimental data because of local variation of magnetic parameters near the edge of the sample due to the e-beam lithography. This is the experimental result that is at the basis of the design of SOT-MRAM based PUF devices. In particular, this configuration can be used to generate a secret key in an array of SOT-MRAMs that is characterized by half of the bit '1' and half '0'.

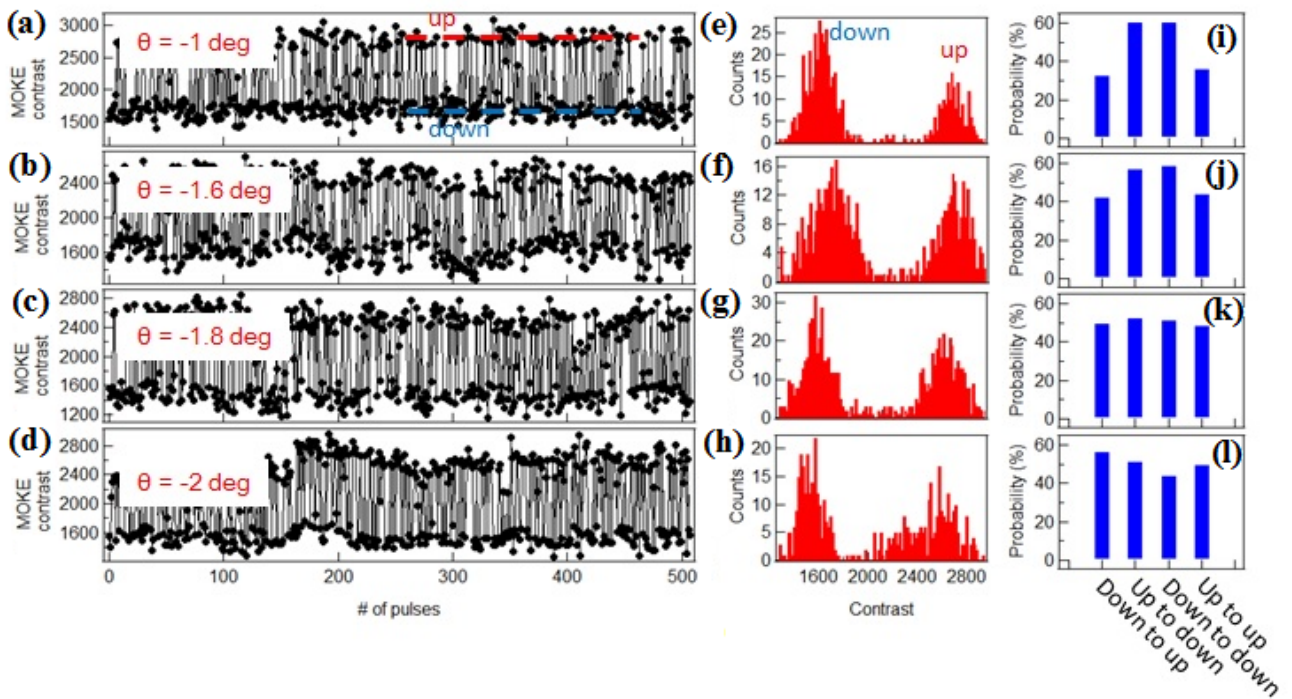

**Figure 3. a-d.** Experimental measurements of the switching of the magnetization, based on the evaluation of the MOKE contrast, for a dot of 1 μm. The switching probability has been computed considering 500 current pulses, each one long 15 ns, and an in-plane field of 40.2 mT applied at different angles as indicated in each panel of these figures. **e-h.** Histograms of MOKE contrast, i.e. probability function of the two out-of-plane directions (down and up) of the magnetization. **i-l.** Switching probability of the 4 possible transitions "up-to-up", "up-to-down", "down-to-up", and "down-to-down" as a response to a current pulse.

## IV Micromagnetic simulations.

To understand qualitatively the FL magnetization dynamics described in the previous paragraph, we have performed micromagnetic simulations carried out by means of a state-of-the-art parallel micromagnetic solver, which numerically integrates the LLG equation including the Slonczewski-like torque due to SHE[17]:

$$\frac{d\mathbf{m}}{d\tau} = -\mathbf{m} \times \mathbf{h}_{EFF} + \alpha_G \mathbf{m} \times \frac{d\mathbf{m}}{d\tau} - \frac{g\mu_B}{2\gamma_0 e M_S^2 t_{CoFe}} \alpha_H \mathbf{m} \times \mathbf{m} \times (\hat{z} \times \mathbf{J})$$ (1)

where $\mathbf{m}$ and $\mathbf{h}_{EFF}$ are the normalized magnetization and the effective field of the ferromagnet, respectively. The effective field includes the standard magnetic field contributions, the Oersted field and the thermal field, which is included as an additional stochastic term.[18] The effective field and boundary conditions expressions are included as in[19],[20]. $\tau = \gamma_0 \cdot M_S \cdot t$ is the dimensionless time, where $\gamma_0$ is the gyromagnetic ratio, and $M_S$ is the saturation magnetization of the ferromagnet. $\alpha_G$ is the Gilbert damping, $g$ is the Landè factor, $\mu_B$ is the Bohr Magneton, $e$ is the electron charge, $t_{CoFe}$ is the thickness of the ferromagnetic layer, $\alpha_H$ is the spin-Hall angle obtained from the ratio between the spin current and the electrical current, $\hat{z}$ is the unit vector of the out-of-plane direction, and $\mathbf{J}$ is the in-plane current density injected via the HM.

The parameters used for the CoFeB are saturation magnetization $M_S = 800 \times 10^3$ Am$^{-1}$, exchange constant A = $2.0 \times 10^{-11}$ Jm$^{-1}$,[21] and damping parameter $\alpha_G = 0.03$. The magnetic anisotropy $K = 5.12 \times 10^5$ Jm$^{-3}$ has been estimated from the experimental hard-axis magnetization curve and the spin-Hall angle $\alpha_H = -0.33$.[22] Fig. 4(a) summarizes the calculations for a field of 40 mT applied with a tilted angle between -1 and +1 degrees and a current of 3.6 mA ($J_{SHE}=6\times10^{-11}$ Am$^{-2}$). First, the simulation results show that the final state probability exhibits the same strong sensitivity to the tilted angle of the field and a fine tuning of this angle is necessary to have a final state probability of 50% as in the experiments (see Fig. 4(b)).

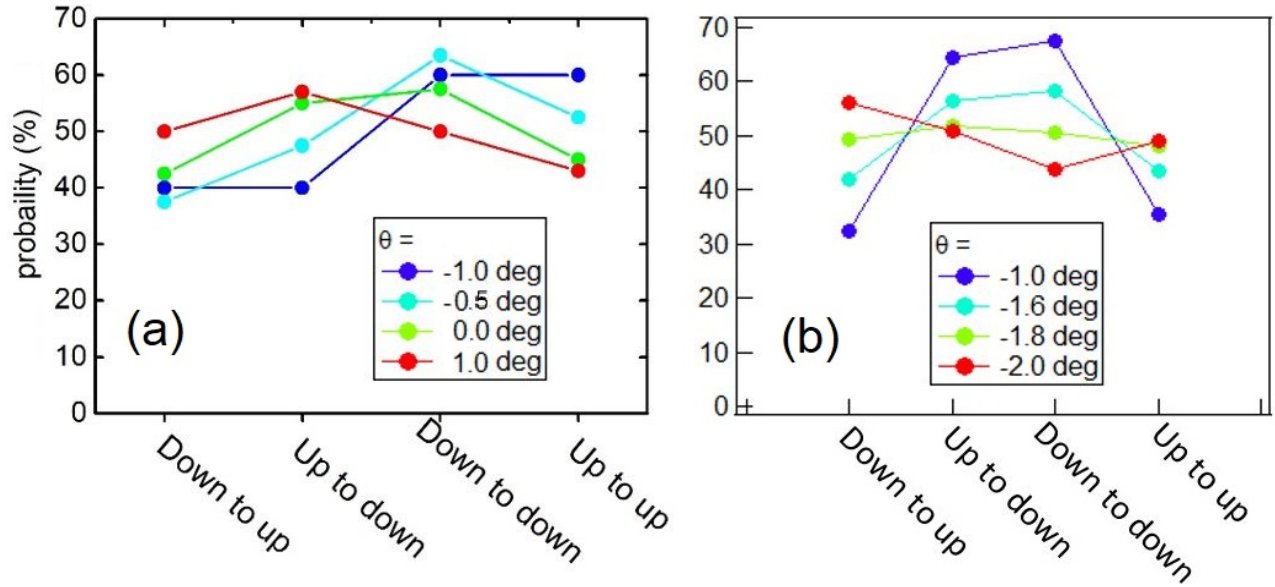

**Figure 4. a.** Results of the micromagnetic simulations and **b.** experimental measurements of the switching probability for the 4 possible transitions "up-to-up", "up-to-down", "down-to-up", and "down-to-down" as calculated at various tilting angles *θ* of the applied magnetic field.

Examples of time domain evolution of the magnetization for the 4 possible transitions (up-up, up-down, down-up and down-down) are shown in Figs. 5(a)-(d). The current pulse drives the initial state (Fig. 5 first column) towards a uniform state of the magnetization along the x-direction that is the same as the direction of the external field (Fig. 5 second column). Once the current is switched off, there is nucleation of domains (Fig. 5 third column) that first expand (Fig. 5 fourth and fifth columns) and then collapse into one single domain state representing the out-of-plane positive or negative configurations (Fig. 5 sixth column).

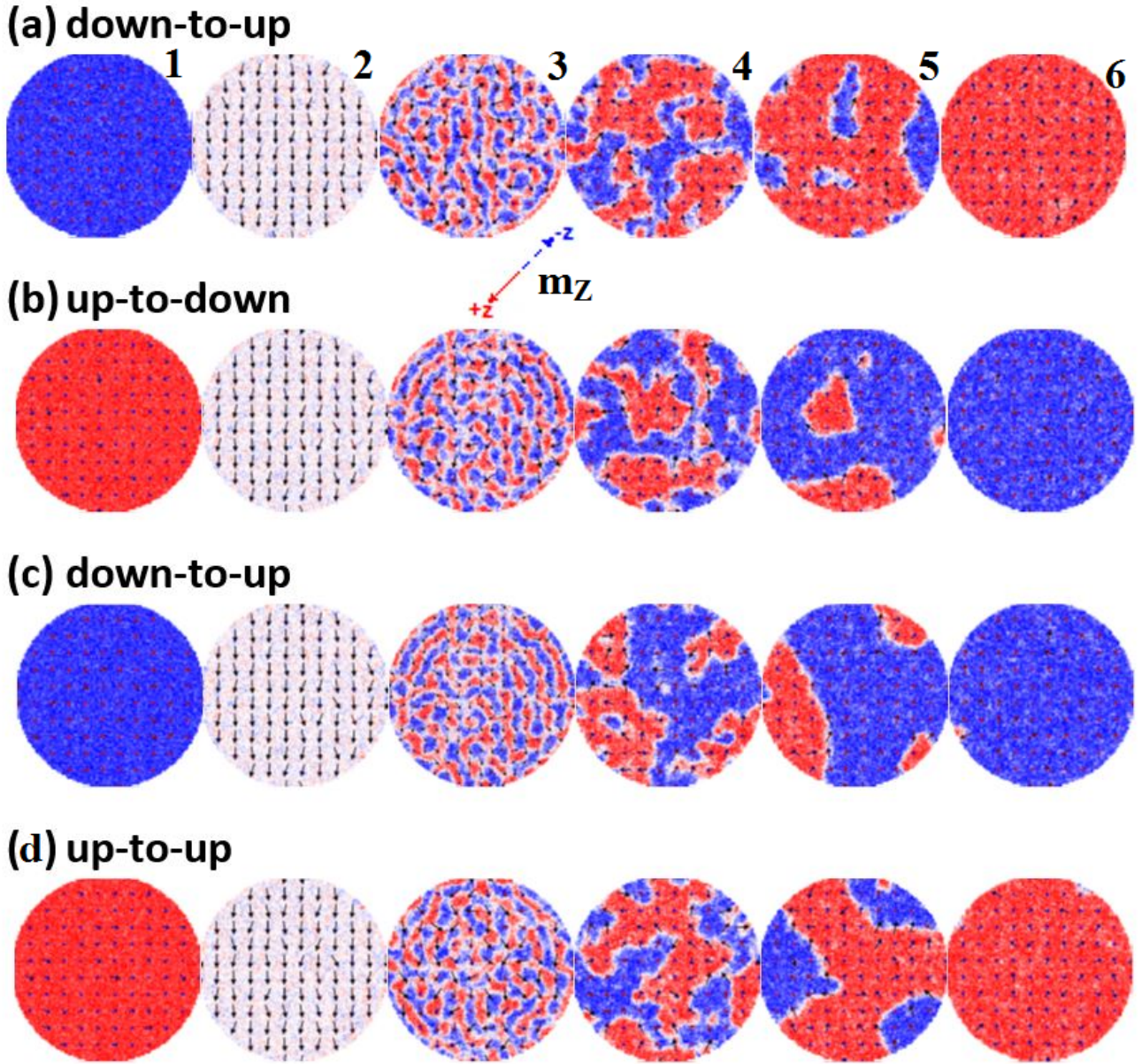

**Figure 5 (a)-(d)** Examples of time domain evolution of the spatial distribution of the magnetization (the colour is related to the out-of-plane component of the magnetization, i.e. red for positive and blue for negative, and the arrows indicate the in-plane component of the magnetization). The rows shows the snapshots for the switching processes for a specific transition **a.** "up-to-up", **b.** "up-to-down", **c.** "down-to-up", and **d.** "down-to-down". The numbering of the columns (number 1-6) is used to characterize the main common steps of the switching process.

## V. Three dots experimental characterization.

To study the robustness of the proposed approach for the design of a PUF, we have built a device composed of three dots within a single nanowire realized geometrically as indicated in Fig. 6(a). The dots have the same size (diameter of 1 μm), while the center-to-center distance is fixed at 5 μm in order to have a negligible effect of magnetostatic coupling. A MOKE image is shown in Fig. 6(b).

We have applied a current pulse of 4 mA (voltage 8 V) for a duration of 15 ns for 100 iterations, while maintaining a field of 40 mT applied at an angle $\theta$ = -0.7 degree. We have found that the switching mechanism is the same for each dot and the same current pulse is able to drive the magnetization of each dot out of the equilibrium qualitatively similar to what observed in single dot measurements. By removing the current pulse, the magnetization of each dot independently goes back to the perpendicular direction, upward or downward, with a certain probability. By using the optical imaging magneto-optical Kerr effect, we have identified the final state of the magnetization after the current pulse is switched off for 100 iterations. In order to process the data, we have modeled specific regions of interest (ROIs) having 1 μm of diameter, i.e. the same as the devices, and centered around the locations of the dots (see the arrows in Fig. 6(b)). We have developed an algorithm which systematically calculates the average brightness of the ROIs against the surrounding background areas and compares it to a probabilistic threshold. The analysis of corresponding cumulative distribution functions allows us to determine the switching probability for each dot with high accuracy. In detail, in those regions, once the current pulse is switched off and the brightness of the regions is evaluated, if the ROI exhibits a different (higher) brightness value against the background, the magnetization is considered upward, and this is linked to the presence of the bit '0'. Conversely, if the region exhibits the same brightness value as in the background, the magnetization points downward and this corresponds to the bit '1'.

The results of this analysis are reported in Fig. 6(c). The extension of the analysis to multiple dots allows us to confirm the intrinsic randomness of the device (ideally it should be close to 50%), which is a crucial property for the unpredictability of the PUF responses.

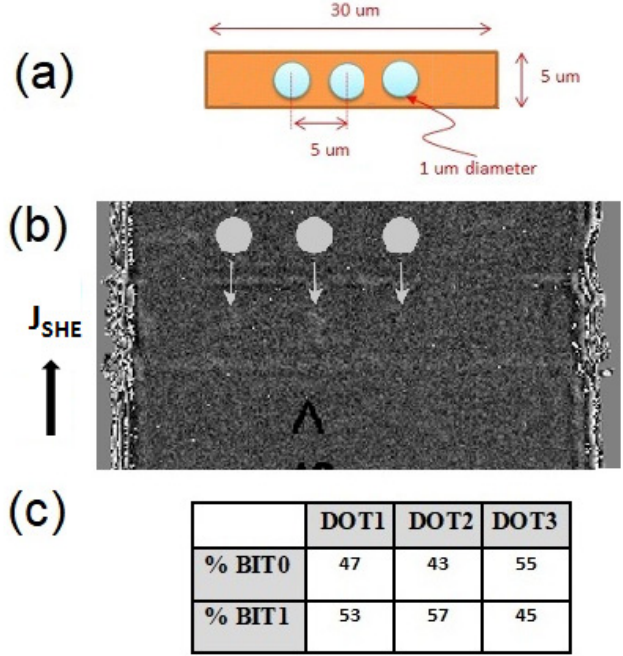

**Figure 6 a.** Schematic of the three dots device. **b.** The optical imaging magneto-optical Kerr effect. **c.** Probability of bit '0' and '1' for each dot during 100 iterations.

## VI. CMOS integration of the SOT-MRAM based PUF.

In the previous sections, we have discussed experimental and simulation analyses performed for magnetic dots with a diameter of 1 μm on top of a W strip with 6 nm in thickness. Here, we evaluate the implications of the proposed PUF approach at the circuit-level in terms of performance and energy characteristics considering that the SOT-MRAM device is built as a point contact geometry with a diameter of 100 nm similarly to a device already developed.[23] In this way, the reading current is applied locally in the FL. To this aim, we have considered a general circuit architecture (see Fig. 7(a)) featuring 4 bitcell (BC) blocks, each including a 16×16 BC array and read/write control circuits, and a decoding block to generate the appropriate read (R) signals for a specific challenge represented by a 64-bit input address (ADD). The latter block consists of 16, one for each row of the BC arrays, 4-to-16 decoders. Each decoder receives 4 bits of the challenge to produce 16 R signals (one for each column of the BC arrays) by a logical AND operation between the decoder output bits and a read/write (RW) control signal. The generated 16×16 R signals are then fed to the 4 BC blocks, whose each one provides 16 output (OUT) bits (one bit for each row), thus obtaining a final 64-bit output word as response to a given challenge.

The scheme of the BC blocks is detailed in Fig. 7(b). The 16×16 BC array relies on 16 strings (one for each row), each one including 16 magnetic dots (one for each column) whose free layers are contacted to the same HM strip in series. The top of each dot is also connected to one NMOS access transistor driven by an R signal to form a BC. All the HM strips are connected on one side to power supply voltage ($V_{DD}$) through one PMOS transistor driven by a write (W) signal and on the other side they are grounded. In addition, the BCs within the same row share a read bitline (RBL). Such a scheme allows managing write and read operations by properly asserting RW, W, and R signals. The write (or program) operation occurs concurrently for all the BCs belonging to one 16×16 array by activating all $V_{DD}$-connected PMOS transistors (all W signals = "0"), while disabling all the NMOS access transistors (all R signals = "0" by setting the RW control signal = "0" in the decoding block). In this way, in each row string, an adequate write current ($I_{write} = I_{SHE}$) flows from $V_{DD}$ to the ground through the HM nanowire, thus enabling the SOT-based switching mechanism in the 16 magnetic dots. This writing operation is performed only one time (or whenever a rewrite of the stored secret keys is required), hence it does not influence significantly the power dissipation of the whole PUF circuit during normal running. Conversely, the read operation occurs as often as a response has to be provided for a specific challenge. This operation is implemented along the RBLs by setting the RW control signal = "1" in the decoding block and hence activating one NMOS access transistor per row in the 16×16 BC arrays (since, for each row, only one R signal coming from the decoding block is equal to "1" on the basis of the given challenge), while disabling all $V_{DD}$-connected PMOS transistors (all W signals = "1"). As shown in Fig. 7(b), for each row, a conventional voltage sensing scheme is then used to detect the binary digit stored in the SOT-MRAM device of the activated BC. In particular, it consists of generating a read current ($I_{read}$) to be applied to the RBL and then to produce a read bitline voltage ($V_{data}$), which is compared with a reference voltage ($V_{ref}$) by a voltage sense amplifier (SA). Obviously, the applied $I_{read}$ has to be sufficiently low to avoid any manipulation of the magnetic state of the SOT-MRAM devices during the reading operation. As a consequence, depending on the FL magnetization orientation (i.e. up or down) in the magnetic device of the activated BC, the developed $V_{data}$ is lower or higher than $V_{ref}$, thus translating into an OUT bit equal to "0" or "1", respectively.

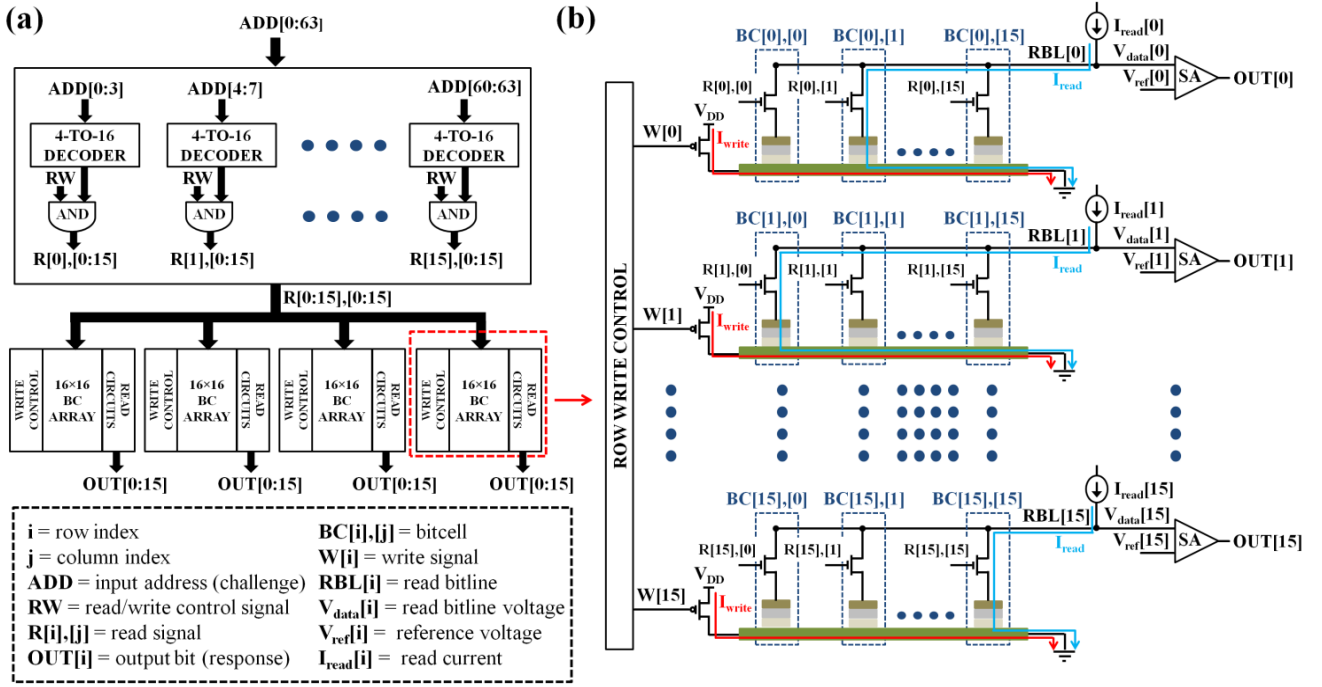

**Figure 7. a.** Block diagram of the CMOS/SOT-MRAM based PUF general architecture along with the description of the signals and **b.** details of the circuit for the bitcell (BC) block.

We have simulated the above hybrid CMOS/SOT-MRAM PUF circuit into Cadence$^{TM}$ environment using the transistor models provided by a 1.8V 0.18 μm CMOS technology and a Verilog-A based compact model[24] to integrate the behavior of SOT-MRAM devices. For the latter, we have considered the following parameters: diameter = 100 nm, $t_{CoFe}$ = 1 nm, MgO oxide thickness $t_{OX}$ = 1.6 nm, resistance-area (RA) product = 1 Ω·μm$^2$, TMR ratio = 150%, $\alpha_G$ = 0.03, $\alpha_H$ = -0.33, $M_S$ = 800×10$^3$ Am$^{-1}$, $W_{HM}$ = 1.5 μm, $L_{HM}$ = 1.5 μm (i.e. the center-to-center distance of two adjacent dots on the same HM strip), HM strip thickness $t_{HM}$ = 6 nm. We have properly sized V$_{DD}$-connected PMOS transistors driven by W signals to achieve an adequate I$_{write}$ pulse amplitude of about 3.6 mA ($J_{SHE}$ = 4×10$^7$ Acm$^{-2}$). This allows one to get the FL magnetization in-plane within 1 ns, thus corresponding to an average write energy per bit (E$_{write\_bit}$) of about 0.39 pJ. We have also sized the NMOS access transistors of the BCs and the read control circuits to obtain a much lower pulse amplitude for the I$_{read}$ (around 10 μA for a pulse duration of 1 ns) in such a way as to ensure a low read disturbance probability during sensing operations. This has led us to estimate an average read energy per bit (E$_{read\_bit}$) of about 0.12 pJ, including the contributions of all peripheral circuitry (i.e. the decoding block, the circuits for the I$_{read}$ generation, and the SAs). When no read and write operations occur (standby phase), the designed circuit exhibits a leakage current of 0.73 mA.

However, thanks to the non-volatility capability offered by the SOT-MRAM devices, the leakage power contribution can be completely cut off by switching off the power supply during the standby phase.

**VII Summary and Conclusions**

This work proposes the realization of an SOT-MRAM based PUF where the generation of the secret key is based on driving a perpendicular magnet into an equilibrium configuration (in-plane direction) by applying a large enough current along with an external field. Once the current is switched off, such in-plane magnetic state becomes unstable giving rise to a random magnetization evolution towards the positive or negative out-of-plane stable configurations. Moreover, we evaluated the circuit level implications in terms of energy requirements by integrating the proposed SOT-MRAM device with a commercial CMOS technology.

Experimental and simulations results reported and discussed in this work prove that this PUF approach allows ensuring a randomness close to 50%. In addition, when compared to state-of-the-art CMOS based PUF implementations, the proposed solution offers several intrinsic advantages owing to the use of SOT-MRAM devices. Among these, higher robustness against voltage and temperature fluctuations (the voltage and temperature dependences of SOT-MRAM devices in both the resistance states are significantly lower than those of the MOSFET conduction), the radiation hardness (MRAM technology being intrinsically rad-hard is already seen as major candidate for radiation-hard memory devices), non-volatile behavior (hence zero standby power consumption), and reconfigurability, which is based on the possibility to implement a refresh mechanism for the secret key via the SOT and the possibility to update its challenge-response table . We wish to highlight that a PUF reconfigurable architecture allows to meet the requirements for some practical applications and can improve the reliability and security of a PUF-based authentication system.[25],[26] Thanks to all these favorable properties along with technological scalability (down to nanoscale nodes to achieve low power consumption while maintaining high thermal stability) and easy integration with CMOS processes, SOT-MRAMs based PUFs could represent a real breakthrough for security applications based on hardware authentication.

**Acknowledgments**

This work was supported under the Grant 2019-1-U.0. ("Diodi spintronici rad-hard ad elevata sensitività - DIOSPIN") funded by the Italian Space Agency (ASI) within the call "Nuove idee per la componentistica spaziale del futuro".,, by JSPS KAKENHI (Grant Numbers 19K21972 and 17H04924), and the Future Development Funding Program of Kyoto University Research Coordination Alliance. This work has been also supported by PETASPIN association.

**References**

[1] U. Rührmair, S. Devadas, and F. Koushanfar, Security Based on Physical Unclonability and Disorder, in *Introd. to Hardw. Secur. Trust* (Springer New York, New York, NY, 2012), pp. 65–102.

[2] C. Herder, M.D. Yu, F. Koushanfar, and S. Devadas, Physical unclonable functions and applications: A tutorial, Proc. IEEE **102**, 1126 (2014).

[3] U. Rührmair, H. Busch, and S. Katzenbeisser, Strong PUFs: Models, Constructions, and Security Proofs, in (Springer, Berlin, Heidelberg, 2010), pp. 79–96.

[4] O. Golonzka, J.-G. Alzate, U. Arslan, M. Bohr, P. Bai, J. Brockman, B. Buford, C. Connor, N. Das, B. Doyle, T. Ghani, F. Hamzaoglu, P. Heil, P. Hentges, R. Jahan, D. Kencke, B. Lin, M. Lu, M. Mainuddin, M. Meterelliyoz, P. Nguyen, D. Nikonov, K. O'brien, J.. Donnell, K. Oguz, D. Ouellette, J. Park, J. Pellegren, C. Puls, P. Quintero, T. Rahman, A. Romang, M. Sekhar, A. Selarka, M. Seth, A.J. Smith, A.K. Smith, L. Wei, C. Wiegand, Z. Zhang, and K. Fischer, MRAM as Embedded Non-Volatile Memory Solution for 22FFL FinFET Technology, in *2018 IEEE Int. Electron Devices Meet.* (IEEE, 2018), pp. 18.1.1-18.1.4.

[5] A. Raychowdhury, MRAM and FinFETs team up, Nat. Electron. **1**, 618 (2018).

[6] S. Khaleghi, P. Vinella, S. Banerjee, and W. Rao, An STT-MRAM based strong PUF, in *Proc. 2016 IEEE/ACM Int. Symp. Nanoscale Archit. NANOARCH 2016* (Institute of Electrical and Electronics Engineers; Association for Computing Machinery; IEEE Computer Society, Beijing, 2016), pp. 129–134.

[7] A. Iyengar, S. Ghosh, K. Ramclam, J. Jang, and C. Lin, Spintronic PUFs for Security, Trust, and Authentication, ACM J. Emerg. Technol. Comput. Syst. **13**, 1 (4) (2016).

[8] T. Marukame, T. Tanamoto, and Y. Mitani, Extracting Physically Unclonable Function From Spin Transfer Switching Characteristics in Magnetic Tunnel Junctions, IEEE Trans. Magn. **50**,

3402004 (2014).

[9] N. Perrissin, S. Lequeux, N. Strelkov, A. Chavent, L. Vila, L.D. Buda-Prejbeanu, S. Auffret, R.C. Sousa, I.L. Prejbeanu, and B. Dieny, A highly thermally stable sub-20 nm magnetic random-access memory based on perpendicular shape anisotropy, Nanoscale **10**, 12187 (2018).

[10] S. Ikeda, K. Miura, H. Yamamoto, K. Mizunuma, H.D. Gan, M. Endo, S. Kanai, J. Hayakawa, F. Matsukura, and H. Ohno, A perpendicular-anisotropy CoFeB-MgO magnetic tunnel junction., Nat. Mater. **9**, 721 (2010).

[11] I.M. Miron, K. Garello, G. Gaudin, P.-J. Zermatten, M. V Costache, S. Auffret, S. Bandiera, B. Rodmacq, A. Schuhl, and P. Gambardella, Perpendicular switching of a single ferromagnetic layer induced by in-plane current injection., Nature **476**, 189 (2011).

[12] L. Liu, C.-F. Pai, Y. Li, H.W. Tseng, D.C. Ralph, and R.A. Buhrman, Spin-Torque Switching with the Giant Spin Hall Effect of Tantalum, Science **336**, 555 (2012).

[13] G. Finocchio, M. Carpentieri, E. Martinez, and B. Azzerboni, Switching of a single ferromagnetic layer driven by spin Hall effect, Appl. Phys. Lett. **102**, 212410 (2013).

[14] D. Bhowmik, L. You, and S. Salahuddin, Spin Hall effect clocking of nanomagnetic logic without a magnetic field, Nat. Nanotechnol. **9**, 59 (2014).

[15] V. Puliafito, A. Giordano, B. Azzerboni, and G. Finocchio, Nanomagnetic logic with non-uniform states of clocking, J. Phys. D. Appl. Phys. **49**, 145001 (2016).

[16] R. De Rose, M. Lanuzza, F. Crupi, G. Siracusano, R. Tomasello, G. Finocchio, and M. Carpentieri, Variability-Aware Analysis of Hybrid MTJ/CMOS Circuits by a Micromagnetic-Based Simulation Framework, IEEE Trans. Nanotechnol. **16**, 160-168 (2016).

[17] J.C. Slonczewski, Current-driven excitation of magnetic multilayers, J. Magn. Magn. Mater. **159**, L1 (1996).

[18] G. Finocchio, I.N. Krivorotov, X. Cheng, L. Torres, and B. Azzerboni, Micromagnetic understanding of stochastic resonance driven by spin-transfer-torque, Phys. Rev. B **83**, 134402 (2011).

[19] A. Giordano, G. Finocchio, L. Torres, M. Carpentieri, and B. Azzerboni, Semi-implicit integration scheme for Landau-Lifshitz-Gilbert-Slonczewski equation, J. Appl. Phys. **111**, 07D112

(2012).

[20] G. Siracusano, R. Tomasello, A. Giordano, V. Puliafito, B. Azzerboni, O. Ozatay, M. Carpentieri, and G. Finocchio, Magnetic Radial Vortex Stabilization and Efficient Manipulation Driven by the Dzyaloshinskii-Moriya Interaction and Spin-Transfer Torque, Phys. Rev. Lett. **117**, 087204 (2016).

[21] L. Zhang, B. Fang, J. Cai, M. Carpentieri, V. Puliafito, F. Garescì, P.K. Amiri, G. Finocchio, and Z. Zeng, Ultrahigh detection sensitivity exceeding $10^5$ V/W in spin-torque diode, Appl. Phys. Lett. **113**, 102401 (2018).

[22] C.-F. Pai, L. Liu, Y. Li, H.W. Tseng, D.C. Ralph, and R.A. Buhrman, Spin transfer torque devices utilizing the giant spin Hall effect of tungsten, Appl. Phys. Lett. **101**, 122404 (2012).

[23] A. Houshang, R. Khymyn, H. Fulara, A. Gangwar, M. Haidar, S.R. Etesami, R. Ferreira, P.P. Freitas, M. Dvornik, R.K. Dumas, and J. Åkerman, Spin transfer torque driven higher-order propagating spin waves in nano-contact magnetic tunnel junctions, Nat. Commun. **9**, 4374 (2018).

[24] Z. Wang, W. Zhao, E. Deng, J.-O. Klein, and C. Chappert, Perpendicular-anisotropy magnetic tunnel junction switched by spin-Hall-assisted spin-transfer torque, J. Phys. D. Appl. Phys. **48**, 065001 (2015).

[25] Le Zhang, Zhi Hui Kong, Chip-Hong Chang, A. Cabrini, and G. Torelli, Exploiting Process Variations and Programming Sensitivity of Phase Change Memory for Reconfigurable Physical Unclonable Functions, IEEE Trans. Inf. Forensics Secur. **9**, 921 (2014).

[26] A. Chen, Utilizing the Variability of Resistive Random Access Memory to Implement Reconfigurable Physical Unclonable Functions, IEEE Electron Device Lett. **36**, 138 (2015).